# Deep learning-based denoising streamed from mobile phones improves speech-in-noise understanding for hearing aid users

Peter Udo Diehl[1,2,*], Hannes Zilly[1], Felix Sattler[1], Yosef Singer[1], Kevin Kepp[1], Mark Berry[1], Henning Hasemann[1], Marlene Zippel[3], Müge Kaya[4], Paul Meyer-Rachner[1], Annett Pudszuhn[2], Veit M. Hofmann[2], Matthias Vormann[4], Elias Sprengel[1]

[1]Audatic, Friedrichstr. 210, 10969 Berlin, Germany

[2]Charité – Universitätsmedizin Berlin, corporate member of Freie Universität Berlin, Humboldt-Universität zu Berlin, and Berlin Institute of Health, Department of Otorhinolaryngology, Head and Neck Surgery, Campus Benjamin Franklin, Germany

[3]Sonova AG, Laubisrütistrasse 28, 8712 Stäfa, Switzerland

[4]Hörzentrum Oldenburg GmbH, Marie-Curie-Straße 2, 26129 Oldenburg, Germany

**\* Correspondence:**
Peter Udo Diehl
peter.u.diehl@gmail.com

**Abstract**

The hearing loss of almost half a billion people is commonly treated with hearing aids. However, current hearing aids often do not work well in real-world noisy environments. We present a deep learning based denoising system that runs in real time on iPhone 7 and Samsung Galaxy S10 (25ms algorithmic latency). The denoised audio is streamed to the hearing aid, resulting in a total delay of around 75ms. In tests with hearing aid users having moderate to severe hearing loss, our denoising system improves audio across three tests: 1) listening for subjective audio ratings, 2) listening for objective speech intelligibility, and 3) live conversations in a noisy environment for subjective ratings. Subjective ratings increase by more than 40%, for both the listening test and the live conversation compared to a fitted hearing aid as a baseline. Speech reception thresholds, measuring speech understanding in noise, improve by 1.6 dB SRT. Ours is the first denoising system that is implemented on a mobile device, streamed directly to users' hearing aids using only a single channel as audio input while improving user satisfaction on all tested aspects, including speech intelligibility. This includes overall preference of the denoised and streamed signal over the hearing aid, thereby accepting the higher latency for the significant improvement in speech understanding.

## 1   Introduction

Approximately 5% of the world population currently suffers from hearing loss, with associated side-effects ranging from social isolation, dementia, depression, cortical thinning and increased mortality (Fisher et al., 2014; Cunningham and Tucci, 2017; Ha et al., 2020). Hearing aids and cochlear implants have been shown to mitigate many of these negative effects. Nevertheless, a persistent complaint of hearing aid users is that current devices do not work well in noisy environments (Hartley et al., 2010; Hougaard and Ruf, 2011). One solution is to improve the signal-to-noise ratio of the sound that is output to the user by employing denoising systems on the device. This has proven difficult to do effectively given the limited processing power available on these devices, and previous filterbank-based denoising systems on hearing aids have not been shown to offer improvements in speech intelligibility in noisy environments without depending on spatial knowledge of the scene (Boymans and Dreschler, 2000; Alcántara et al., 2003; Mueller et al., 2006; Zakis et al., 2009; Brons et al., 2014; Völker et al., 2015; Chong and Jenstad, 2018).

The rise of deep learning and especially its increased use in audio, e.g. for speech recognition and speech synthesis, offers a new approach to denoising audio. Deep-learning-based systems achieve state of the art denoising performance (Cao et al., 2022; Tzinis et al., 2022), with some systems offering large improvements in intelligibility, while only using a single channel of audio (i.e., without the need for spatial information) (Goehring et al., 2016; Zhao et al., 2018; Healy et al., 2021; Diehl et al., 2022a). The next step is to downscale these systems, which typically have large computational requirements and can only be used offline, so that they can be used in real-time on mobile and portable systems, such as smartphones and hearing aids. Since deep learning systems scale well with number of parameters and therefore in computing power (Tay et al., 2022), this reduction in size typically causes a reduction in the systems' output quality. Ultimately, hearing aids should be able to provide the computational resources to house powerful denoising systems themselves but currently, mobile phones' advanced processors give them an edge in how computationally expensive, and therefore how good, their denoising systems can be. So far no studies using denoising based on single-channel deep learning models on compact mobile systems (phones or hearing aids) have shown any speech intelligibility improvements, although a few studies have shown improvement in offline computational metrics (Panahi et al., 2016; Hansen et al., 2019; Baby et al., 2021). The largest improvements in speech intelligibility that have been shown on mobile systems are around 0.5 dB SRT (Asger Heidemann Andersen et al., 2021) but require multi-channel input.

In this study, we present a deep learning based single channel denoising system, which retrieves clean speech from a noisy mixed signal. The entire system runs on a commercially available smartphone, which streams the resulting (cleaned) audio to a hearing aid or cochlear implant. The system improves speech intelligibility and overall audio quality without using spatial information. We test the system with hearing aid users across three different tests, including 1) rating the subjective denoising quality (overall, noise, intelligibility, speech quality), 2) objective speech intelligibility, and 3) comparing the phone-based denoising to a hearing aid only in a live setup where the subjects have a conversation with the experimenter. In all three tests, our denoising system achieves significant improvements over the baseline (where only the hearing aid, but no denoising system is used) in overall impression and in speech intelligibility.

## 2  Material and Methods

### 2.1  Denoising System

The presented denoising system was implemented on an iPhone 7, utilizing only a single processor core, and on a Samsung Galaxy S10. Our denoising system is identical on both phones; we chose two different phone manufacturers to ensure that despite the differences in the respective microphones and audio processing paths, both are compatible with our denoising system.

From an input short time Fourier transform (Hann window with 25ms length, 6ms hop size, and 22kHz sampling rate) of mixed speech and noise, the network is trained to predict the complex ideal ratio mask (Wang and Chen, 2018). The input samples are single-channel mixtures of speech and noise drawn from multiple publicly available databases with ca. 10,000 hours of audio (Panayotov et al., 2015; Gemmeke et al., 2017; Afouras et al., 2018; Wang et al., 2021); target samples are the speech samples without any noise added.

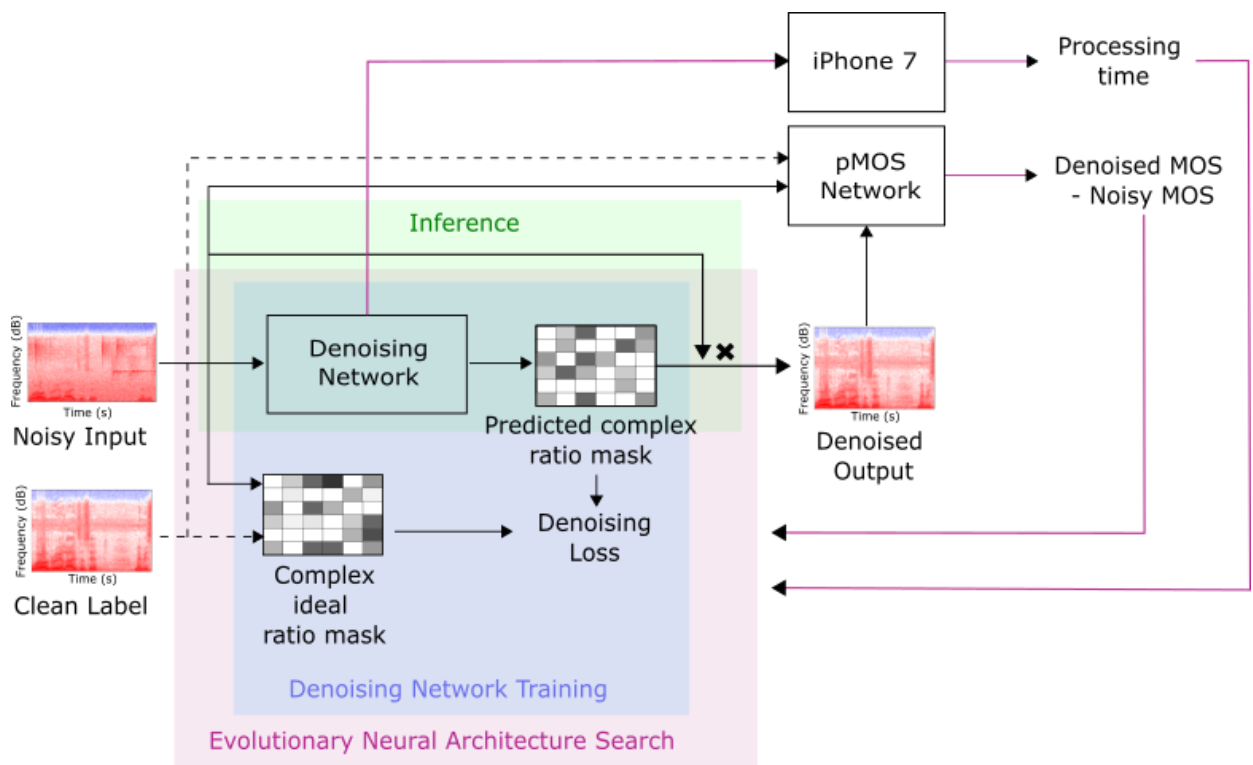

**Figure 1. Training pipeline of the denoising system, based on a mean opinion score (MOS)-estimator-guided neural architecture search. The denoising network predicts denoised outputs, given mixed speech and noise input STFTs. Network parameters, such as number of layers or type of filters, are optimized by an evolutionary neural architecture search. This search minimizes the remaining error for human acoustic perception by using a MOS estimator, which is a deep neural network trained on a dataset generated from over 1,000,000 human rated audio files.**

We use a closed-loop evolutionary search that optimizes the network parameters and its structure via a custom speech quality metric in combination with the measured execution time of the network on an iPhone 7 (Figure 1). The speech metric (predicted mean opinion score, pMOS in Figure 1) predicts mean opinion scores (MOS) and the improvement in predicted MOS scores is used for optimization. The speech metric is publicly available at https://metric.audatic.ai/ (Diehl et al., 2022b). The closed-loop evolutionary search allows the co-optimization of denoising performance together with the limited resources on the mobile device. We limit the maximum execution time to 25ms, while searching for the model with the best possible denoising performance. This upper bound was chosen because preliminary tests had shown processing delays above 25ms were noticeable, but acceptable to internal testers. Latency limitations have strong effects on denoising performance, since higher latencies allow more computation to be done and the possibility to provide more temporal context for the neural networks.

The basic network architecture is inspired by the U-Net (Ronneberger et al., 2015), with parameters such as layer type (convolutional, Long-Short Term Memory (LSTM), Gated Recurrent Units (GRU), Convolutional Recurrent Network (CRNN) etc.), possible skip connection locations, temporal and spectral down- and up-sampling, number of layers, as well as their size left unspecified to be then optimized by the evolutionary search. The resulting best performing network architecture is then retrained using Population Based Training (PBT) (Jaderberg et al., 2017), which jointly optimizes the network weights and hyperparameters such as the learning rate. To further reduce the computational footprint of the network and to prevent thermal throttling during continuous execution on the phone, we apply structured magnitude pruning. More specifically, we iteratively prune a small fraction of low-magnitude output channels in each layer, starting in the final layers of the network and progressing to the input layers. We then fine-tune the network weights through additional training between each pruning operation in order to recover lost performance. This iterative pruning procedure (Li et al., 2017) is then repeated several times. The final model is again fine-tuned using PBT. With this procedure we achieve a reduction in MAC (multiply-accumulate) operations of 22% from the original model chosen through the neural architecture search with 37.9M to 29.5M while only suffering a minor degradation in speech quality performance. Aside from pruning we also experimented with matrix and tensor decomposition techniques (Lebedev et al., 2015; Kim et al., 2016; Kuchaiev and Ginsburg, 2018) which resulted in a similar reduction in the number of model parameters. However, the resulting factorized architectures, where large layers are replaced with a series of smaller layers, did not lead to reduced execution time on the phone hardware.

The amount of filtering performed by the denoising system can be set by the user from 0% to 100% in increments of 1% (101 steps). This value is referred to as the "mixing ratio". The mixing ratio linearly interpolates between the original input signal and the denoised audio generated by the denoising system. Therefore, a mixing value of 0% passes the unchanged input signal to the streamer, while a mixing value of 100% only passes the denoised audio without mixing any of the original noisy input signal back into the output. Mixing allows the subject to reintroduce environmental clues, reduce effects of isolation and increase environmental awareness and can improve perceived sound quality compared to using the fully denoised signal. No additional postprocessing is applied on the audio.

## 2.2  Subjects and Inclusion Criteria

The experimental protocols employed were approved by the ethics committee ("Kommission für Forschungsfolgenabschätzung und Ethik") of the University of Oldenburg, Oldenburg, Germany and concur with the Helsinki Declaration. All subjects gave their informed consent, were selected from the Hörzentrum Oldenburg subject database, and were paid an expense reimbursement of 12€ per hour. Their participation was voluntary.

In total, 26 German speaking hearing impaired subjects (19 male, 7 female) aged between 46 and 85 (median 75) and an average hearing loss of 65 dB HL at 1 kHz (Figure 2) participated. All participants had been using hearing aids for at least three years prior to their participation in the study. Additionally, three pilot-subjects were included before the actual study to ensure feasibility of the tests and procedures. Inclusion criteria for all participants were: 1) to be at least 18 years old, 2) to have moderate to severe sensorineural hearing loss (in the range of hearing profiles between N4 and N6 (Bisgaard et al., 2010)) to be experienced hearing aid users (>3 years experience). Exclusion criteria were 1) suspected dementia, indicated by DemTect (Kalbe et al., 2004) values of 8 and below, since this will likely influence performance on the tasks.

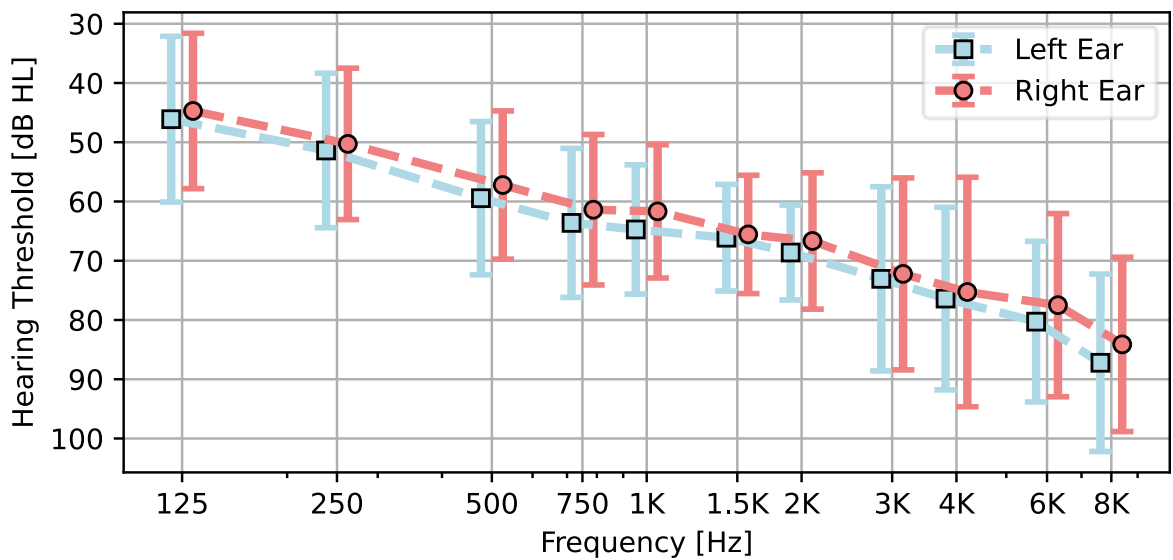

**Figure 2. Mean audiogram of the subjects. Bars show the standard deviation for each frequency. The Phonak Marvel 90 hearing aids were bilaterally fitted to the measured audiogram of each subject.**

### 2.3 Experimental Setup

We tested our system with three different procedures (see Figure 3): A) subjective ratings of a variety of sound scenes, done with offline processing on a PC, enabling a double-blind setting, B) single-blind speech intelligibility tests, with live processing (performed only on the Android phone), C) live conversations with the experimenter, with live processing (performed on both Android and Apple phones). The experiments were conducted over a timespan of three days. Before measurements, all subjects were fitted with hearing aids and customized earmolds (Phonak SlimTip).

The signals recorded with the smartphone microphone were processed by the denoising system and sent from the output device (computer in the sound sample rating, and phone in speech intelligibility and live conversation) to a "streamer", the Phonak TV Connector. Streamers are available as an accessory for modern hearing aid systems produced by all large manufacturers and relay a signal from a source to a hearing aid wirelessly. The hearing aid applies no additional sound processing on relayed signals. The streamer latency was approximately 20ms."

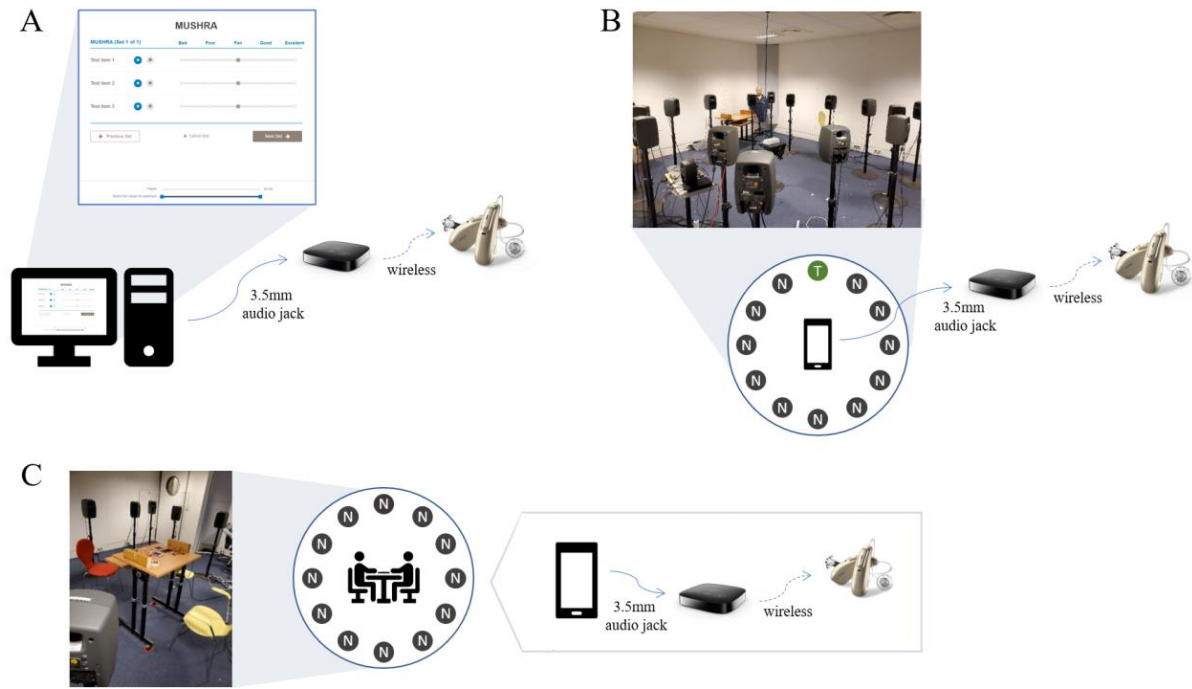

**Figure 3. Testing setups for the 26 hearing-aid user subjects. A) Computer-based subjective listening test (MUSHRA). Ratings of presented audio samples are collected across the categories overall, noise, sound quality, and intelligibility. B) Objective intelligibility is tested using the OLSA. 12 loudspeakers are arranged in a circle with the target loudspeaker presented away from the phone microphone and all 11 other loudspeakers playing noise. C) Live conversations are also held with 12 surrounding loudspeakers, all of which are playing noise. The phone is placed in the middle of the table. In all three setups the audio that was denoised with our system is transmitted to a TV connector which then transmits wirelessly to the hearing aid.**

## 2.4 Subjective ratings using MUSHRA

To assess the subjective improvement using our denoising system in an offline setting, we use a version of Multi-Stimulus test with Hidden Reference and Anchor (MUSHRA) (ITU-R BS.1534-3 Recommendation. Method for the subjective assessment of intermediate quality level of audio systems, 2015) without repeating the reference sound. Specifically, an unprocessed sound ("reference signal") is presented, along with two processed versions of the same signal with different mixing ratios. The subject rates each sample, while being able to switch back and forth between them without interrupting the signal.

During the measurements, participants are seated in front of a computer and asked to rate the samples. The output of the PC is transferred via digital-analog conversion (RME Fireface UC) directly to the streamer which relays it to the hearing aid (Figure 3A). At the beginning, the presentation volume of the MUSHRA test is set to a subjectively comfortable volume for the subject, such that the subject is able to perceive both speech and background noise clearly in all scenes without being affected by too loud noise or too quiet speech components. Ratings are evaluated on a scale from 0 to 100 (101 steps) in the following four categories: 1) Speech Intelligibility, 2) Sound quality of the speech, 3) Background noise, 4) Overall impression.

The sound samples used in this offline test were recorded with an iPhone 7 and then processed by the same denoising system that is implemented on the phones. This ensures a double-blind comparison without adjusting the mixing ratio, neither by the experimenter nor by the subjects. For each sound sample, the unprocessed sample is presented together with the denoised sound at two mixing ratios: 80% and individual preference. The individual mixing ratio was chosen by the subjects during an automated procedure (similar to a binary search) that determines the preferred ratio for each subject with a step size of 5%. The individual mixing ratio is kept constant throughout the experiment. Six different acoustic scenes / sound samples with different signal-to-noise ratios (SNR in the range of approximately -6.6 to 5 dB) were tested: 1) Busy restaurant (low SNR) with a single speaker; 2) Bistro (moderate SNR) with a single speaker; 3) Bistro (moderate SNR) with multiple speakers in the same conversation; 4) Lounge / bar with background music (moderate to high SNR) with a single speaker; 5) Street scene with a single speaker; 6) Street scene with multiple speakers in the same conversation. This results in a total of 24 sound samples (18 processed by the denoising system and 6 unprocessed). The six scenes are presented in random order, for each of the four rating categories, which have a fixed order (Speech Intelligibility, Sound quality of the speech, Background noise, and then Overall impression).

## 2.5 Measuring objective speech intelligibility using the Oldenburg Sentence Test (OLSA)

For objective measurement of speech intelligibility in noise, the Oldenburg Sentence Test (OLSA) was used (Wagener et al., 1999; Wagener, K. C., Brand, T., Kollmeier, B., 1999). Participants repeat as many words as possible from a 5-word sentence presented to them in the presence of background noise. During the experiment, the noise level remains constant whereas the speech level is adaptively adjusted in a two-up/one-down procedure such that the subject is able to understand approximately 70% of the words.

The final SNR of the speech compared to the noise is referred to as the Speech Reception Threshold (SRT). To reduce training effects during measurements, two lists of sentences (20 sentences each) are presented before the actual test. One of the two training lists is presented with a mixing ratio of 0%, the other with 80% with order balanced across participants. The loudness of the streamer is calibrated during the first list, such that the noise is clearly perceptible, but the overall presentation is not perceived as uncomfortably loud. Subsequent measurements are completed with this setting. Here, the noise consisted of many-talker babble noise, recorded in a crowded cafeteria, with an overall level of 65 dBA.

For the OLSA measurements, the system runs on a Samsung Galaxy S10 smartphone, which sends the processed signals to the streamer, that forwards the signals to the hearing aids of the subject, who sits in an adjacent room (Figure 3B). The noise is played from 12 loudspeakers that are identically spaced around the smartphone on which the denoising system is running. The loudspeaker in front of the phone, i.e., opposite of the smartphone's microphone, plays the OLSA sentences. To activate the streamer, a sentence is played that announces the beginning of the listening test.

Tests are performed in four different mixing ratios: 0%, 50%, 80%, and individual preference. The individual mixing ratio for this test is determined by playing OLSA sentences in noise and asking the subject to determine their favorite setting. The measurements are conducted in a balanced order (latin square design) of mixing ratios. A break of at least 10 minutes is taken following the initial procedure.

## 2.6 Live conversation subjective ratings

The goal of the live conversation setup is to simulate a typical situation at a busy cafeteria and compare the subjects' satisfaction using the hearing aid and using the denoising system (which includes the latency caused by the wireless link). To this end, noise is presented from 12 equally spaced loudspeakers, centered around a table where the subject and the experimenter sit, facing each other at a distance of about 1.25 m (Figure 3C). The 12 loudspeakers play a bistro scene at a total level of 68 dBA, that is comprised of noises like babbling noise, water faucets, and sounds of clearing plates. Due to hygiene concerns related to the COVID pandemic, a shield was placed between the subject and experimenter with a window at the bottom (similar to shields used at cashiers). During the conversation, the experimenter tried to maintain an equal duration of speaking himself and letting the subject speak to allow them to judge the sound of their own voice, too. This is important as own voice perception is the most likely to be impacted by the latency of the phones (ca. 10ms on iPhone 7; 20ms on Samsung Galaxy S10), streamer (ca. 10ms), wireless link (ca. 20ms), and the algorithmic latency of ca. 25ms.

The live conversation starts off with normal usage of the hearing aid without streaming (contrary to MUSHRA and OLSA). After 5 to 10 minutes of conversation, the streamer and denoising system are activated with the mixing ratio set to 0% (no processing) so the subject can calibrate the loudness such that it matches the hearing aid. This adjustment is done for both smartphones and maintained for all subsequent measurements. The experimenter and the subject conduct a dialogue for at least 5 minutes and then rate each of the following four categories: overall preference, own voice, experimenter's voice, and noise reduction. The subject selects ratings on a scale from 1 to 100 using an app on a tablet (Figure 3C, bottom). Finally, the subjects' task is to compare the subjective quality of the hearing aid vs. denoising on the iPhone 7 vs. denoising on the Samsung Galaxy S10. Both denoising implementations on both phones work identically and are set to a mixing ratio of 80% but use the respective audio processing stack of the phones (i.e. the operating system dependent pre- and post-processing of the audio, since there is no direct access to the raw audio without going through the operating systems audio framework). The hearing aids used the speech in noise program (default setting).

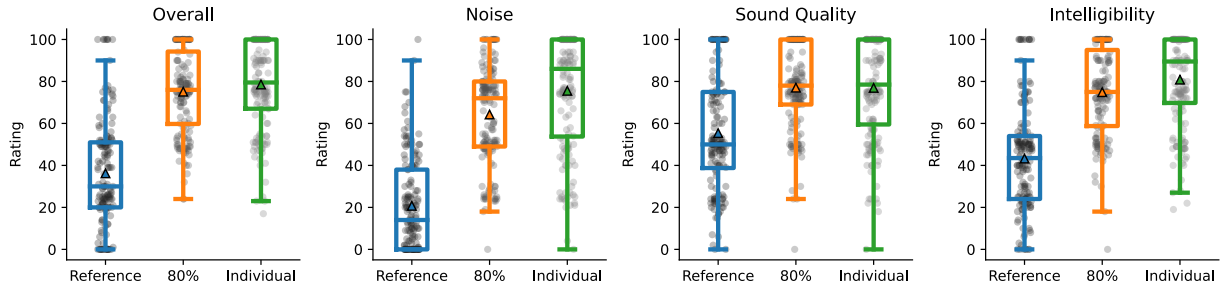

**Figure 4. MUSHRA results of hearing-impaired subjects comparing the reference signal to the deep learning processed signal with self-adjusted mixing ratio (mean mixing ratio is 81.15%, n=26). Subjective ratings on the categories overall, noise, sound quality, and intelligibility. Bars indicate 75, 50, and 25 percentiles. Triangles represent the mean.**

## 3 Results

We first tested the subjective impact of our denoising system (Figure 4). The ratings from the MUSHRA test show a strong improvement for the overall and noise categories when using our denoising system, with 42.3 points and 54.7 points of mean improvement respectively, on a 1 to 100 scale, using individual mixing ratios. Subjects are presented the 1 to 100 scale in words to express how much they like a sample (German: "Gefällt mir") with 5 anchors: 100 – very good ("sehr gut"), 75 - good ("gut"), 50 - neutral ("weder noch"), 25 - dislike ("nicht"), 0 - strongly dislike ("gar nicht"). There is also an improvement of 21.5 points in perceived sound quality with our denoising system (for the individual mixing ratio setting). The perceived intelligibility also improved by 37.6 points using the individual mixing ratio. The individual mixing ratio of the MUSHRA (mean 81.15% processed, median 100%, range from 0% to 100%, STD of 27.6%) is kept the same over the four different rating categories. The improvement in overall rating due to our denoising is independent of the users' hearing loss as measured by PTA Combined (r = -0.02 for individual mixing ratio, r = -0.003 for 80% mixing ratio).

Another key aspect of hearing aids is their ability to provide the user with an objectively measurable increase in speech understanding. We test for this using the OLSA speech in noise intelligibility test (Figure 5). For the OLSA, the mean individual mixing ratio was 71.25% with a standard deviation between subjects of 20.27%. The reduction of the mixing ratio compared to the MUSHRA test is likely because the OLSA operates at lower SNRs (up to -8dB) compared to the MUSHRA, which we performed in the range of -5 to 5 dB SNR. At such low SNRs, the denoising system produces more artifacts than at higher SNRs. Without any denoising, the average SRT for the 26 subjects is at -4.3 dB (std. 1.69 dB) SRT (Figure 5A). When using our denoising system, this improved to -5.45 (std. 1.39 dB) SRT, -5.51 (std. 1.22 dB) SRT, and -5.53 (std. 1.29 dB) SRT with mixing ratios of 50%, 80%, and individual preference, respectively. The resulting increases in SRTs are **1.14 (std 0.9 dB) SRT, 1.20 (std. 1.08dB) SRT, and 1.22 (std. 0.83 dB) SRT (Figure 5B).** When visiting the hearing care professional who fits the hearing aid, it is possible to choose the most beneficial setting for each hearing aid user. Similarly, when choosing the best mixing ratio for each subject (excluding a mixing ratio of 0%, since none of the results were optimal without using our denoising), it improves to a mean of -5.95 (std. 1.17 dB SRT), thereby enabling the subjects to maintain their speech understanding at a 1.64 dB lower speech level than without denoising (Figure 5B). Furthermore, choosing the best mixing ratio, the system always increases the SRT without any negative SRT changes among the 26 subjects. Additionally, we observe a correlation between initial SRT and SRT improvement with our denoising system of r=0.62 (Figure 5C). This indicates that subjects with more severe hearing loss (as measured by the unprocessed SRT), benefited more strongly from our denoising system.

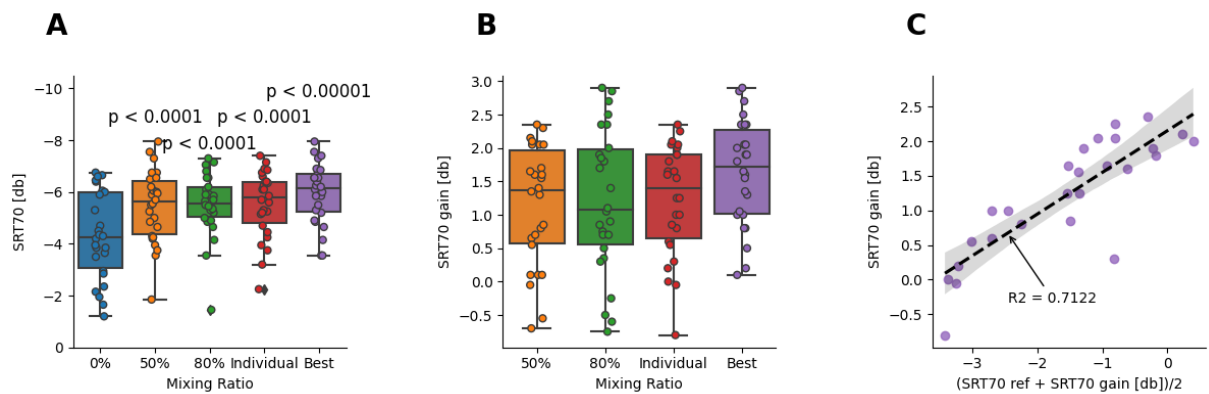

**Figure 5. Reducing speech reception thresholds (SRT) on the Oldenburg Sentence Test using our denoising system for hearing impaired (n=26) subjects at 70% speech intelligibility (SRT70). A) Distribution of SRT70 for different mixing ratios (0%, 50%, 80%, ind.). P-values are based on the t-test between 0% mixing ratio and the displayed mixing ratios. B) Improvements over 0% mixing ratio for all other mixing ratios. C) Relationship between SRT gains in the best mixing ratio setting (besides 0%) and baseline SRTs without denoising (0% mixing), adjusted to avoid spurious correlations.**

The denoising system or similar versions are intended to be used in real-life scenarios and as such, tests in front of the computer or via pre-recorded and standardized sentences do not fully reflect how these systems perform in the real world. To bridge the gap between a laboratory setting and a real-life scenario, we performed a live conversation experiment where we asked the participants for their subjective preference between a standard hearing aid and our denoising systems on the two phones (Figure 6). The implementations of our denoising system on the two phones did not differ algorithmically and showed similar improvements of 38 / 43 points better on average for the overall category than the hearing aid on a 1-100 scale (Figure 6), for the iPhone 7 ("iOS") and the Samsung Galaxy S10 ("Android"), respectively.

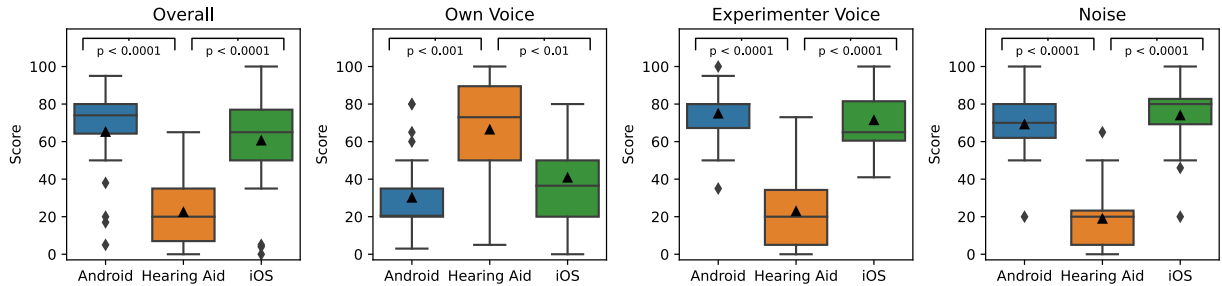

**Figure 6. Live conversation ratings for hearing impaired subjects (n=26) comparing our denoising system on an iPhone 7 (iOS), a Samsung Galaxy S10 (Android), and a fitted hearing aid. Responses for overall ratings are asked first, own voice, experimenter voice and noise are asked in that order afterwards.**

The overall impression of the system includes multiple factors. We therefore also asked subjects for their ratings in three other categories: own voice, experimenter voice, and noise (Figure 6). While the experimenters voice and noise ratings are in line with the overall ratings, the own voice ratings are worse than the hearing aid when using the phone-based denoising system. This is likely because there is a high delay caused by the wireless link and audio processing stack on the phone (in total more than 70 ms of which the iPhone contributes ca. 10ms and Android contributes ca. 20ms). When we asked subjects for the reasons of their low own-voice ratings many reported an "echo" in their voice. Given that the scene was chosen to be a challenging acoustic environment with the experimenters intentionally maintaining a normal conversational speech-level without raising their voice, subjects still showed a strong overall preference for the denoising model despite the delay and poor own-voice perception because they could better understand the experimenter. We do not expect this preference to be maintained in situations with little-to-no background noise.

## 4 Discussion

We tested a deep-learning based denoising system, implemented on two different phone platforms, with 26 hearing aid users in three different test setups. In the MUSHRA test, overall impression ratings increased by 42 points (on a 0 to 100 scale); in the OLSA test the speech reception threshold improved by 1.6 dB; and in the live conversation, overall impression compared to a hearing aid improved by 54.7 points (on a 0 to 100 scale). Using this comprehensive test suite, we are the first to show improvements across all tests (and especially speech intelligibility), using a single-channel denoising system.

In the MUSHRA test, noise ratings improve the most, likely driving the strong improvement in the overall rating category. The OLSA test showed that our system provides higher intelligibility improvement for individuals with worse baseline intelligibility. Additionally, the variance of the intelligibility is reduced when choosing a mixing ratio of at least 80%. This is mostly achieved by improving the worst speech intelligibility results, while moderately increasing the rest. The OLSA test also revealed that it might be useful to adjust the mixing ratio in a real-world environment depending on the estimated SNR. Lower mixing ratios (e.g. the 73% average individual mixing ratio preferred by the subjects in the OLSA test) might lead to better intelligibility in extreme situations (-5 dB to -10 dB) and higher mixing ratios (80%+ preferred by the users in the MUSHRA test) improve the noise reduction and comfort. Finally, the live conversation experiment shows that the system is preferred to existing top-of-the line hearing aids in the given high-noise situations. The main issue remaining is latency due to the streaming of the signal from the phone to the hearing aid, which is reflected in the own-voice ratings.

One of the advantages of our presented system is that it can be combined with spatial algorithms when a similar system is implemented on the hearing aid or cochlear implant directly. It can be used in a modular way and be activated in the most challenging situations, where powerful neural networks are required, to conserve power and only rely on traditional denoising methods when powerful denoising is not required.

In summary, the presented system already improves upon current hearing aids in high-noise situations and benefits the user across multiple dimensions that include objective intelligibility and subjective preference. The biggest current shortcoming of the system is the latency caused by using a Bluetooth connection. This could be solved by either further scaling the system down and implementing it directly on the hearing aid or by using faster protocols such as aptX by Qualcomm or using Apple's *Made for iPhone* audio processing stack. Of these two options an implementation directly on the hearing aid is preferable because even the fastest communication protocols typically add at least 20ms latency to the system, in addition to (at least) 10ms latency caused by the audio processing stack of the phone. Such a hearing aid implementation would further improve the user experience in real-world usage. However, despite the noticeable levels of latency and the obvious need for improvement, in highly noisy environments the presented solution is still clearly preferred and leads to better speech understanding compared to using the existing hearing aids.

## 5 Conflict of Interest

The authors declare that the research was conducted in the absence of any commercial or financial relationships that could be construed as a potential conflict of interest.

## 6 Author Contributions

P.U.D. designed the concept for this work. H.Z., M.B., P.M.-R., K.K., Y.S., H.H., and F.S. created the denoising system and phone implementation. E.S. and P.U.D. supervised the creation of the denoising system and implementation. P.U.D, M.V, M.Z, and M.K. designed the studies. M.V. supervised the in-person studies with hearing impaired and normal hearing. H.Z., M.V., and F.S. performed the statistical analysis. M.K. conducted the in-person studies. P.U.D, Y.S., H.Z., F.S, designed the figures and illustrations. P.U.D, V.M.H., A.P., Y.S., H.Z., M.V, P.M.-R., and E.S. contributed to the interpretation of the results. P.U.D, H.Z., F.S., Y.S., K.K., M.B, H.H., M.Z., P.M.-R., A.P., V.M.H., M.V., E.S. wrote the manuscript. All authors provided critical feedback and helped revise the manuscript.

## 7 Acknowledgments

We thank Matthias Latzel, Volker Kühnel, Hannes Wüthrich, and Sascha Liebe for fruitful discussions during the conception and development phase and helpful ongoing exchanges about the project.

# 9 Data Availability

The metric used to evaluate the network is publicly available for use at https://metric.audatic.ai/**.**